# Double polarization hysteresis loop induced by the domain pinning by defect dipoles in HoMnO$_3$ epitaxial thin films

D. Lee,[1] H. S. Kim,[2] S. Y. Jang,[1] K. W. Joh,[3] T. W. Noh,[1] J. Yu,[2] C. E. Lee,[3] and J.-G. Yoon[4,*]

[1]*ReCOE and FPRD, Department of Physics and Astronomy, Seoul National University, Seoul 151-747, Republic of Korea*

[2]*CSCMR, Department of Physics and Astronomy, Seoul National University, Seoul 151-747, Republic of Korea*

[3]*Spin Dynamics Lab, Department of Physics, Korea University, Seoul 136-701, Republic of Korea*

[4]*Department of Physics, University of Suwon, Kyunggi-do 445-743, Republic of Korea*

We report on antiferroelectriclike double polarization hysteresis loops in multiferroic HoMnO$_3$ thin films below the ferroelectric Curie temperature. This intriguing phenomenon is attributed to the domain pinning by defect dipoles which were introduced unintentionally during film growth process. Electron paramagnetic resonance suggests the existence of Fe$^{1+}$ defects in thin films and first principles calculations reveal that the defect dipoles would be composed of oxygen vacancy and Fe$^{1+}$ defect. We discuss migration of charged point defects during film growth process and formation of defect dipoles along ferroelectric polarization direction, based on the site preference of point defects. Due to a high-temperature low-symmetry structure of HoMnO$_3$, aging is not required to form the defect dipoles in contrast to other ferroelectrics (e.g., BaTiO$_3$).

Recently, multiferroic (MF) materials have attracted a great deal of attention for their potential applications using the coupling between electrical[1,2] and magnetic properties.[3,4] Until now, the known coupling effects in most intrinsic MF materials were weak and observed mostly at low temperatures. This has motivated researchers to systematically tune the physical properties of the known MF materials, or to exploit artificial MF materials. Several groups have used the epitaxial stabilization technique, which employs a coherent interface strain between thin film and substrate, as a possible route to the exploitation of MF materials. For example, we fabricated hexagonal $R$MnO$_3$ ($R$ = Gd, Tb, and Dy) epitaxial thin films and showed that their MF properties could be enhanced compared to those of their bulk perovskite counterparts.[5–7]

Despite extensive efforts, numerous physical properties of the MF materials remain unclear. The artificially synthesized hexagonal $R$MnO$_3$ epitaxial thin films showed unexpected antiferroelectric (AFE)-like double polarization hysteresis (*P-E*) loops below room temperature.[5–7] Moreover, hexagonal HoMnO$_3$ (HMO) epitaxial thin films, whose bulk crystal structure is hexagonal, also showed the unusual AFE-like double *P-E* loops,[8] as shown in Fig. 1(a). This is rather unexpected because bulk HMO has an AFE phase at high temperatures between 875 and 1300 K and a ferroelectric phase below $T_C$=875 K.[9] However, the origin for the double *P-E* loops has not been yet elucidated but inferred to be due to such effects as epitaxial strain, domain pinning, etc.

Here, we discuss the physical origin of the double *P-E* loops observed in HMO thin films by performing electron paramagnetic resonance (EPR) measurements and first principles calculations. Our results support that the double *P-E* loops could be induced by domain pinning by defect dipoles ($D_{\text{defect}}$). Compared to other ferroelectrics, which usually have a non-distorted, centrosymmetric structure at high temperatures (>700 K), HMO has a highly distorted low-symmetry ($P6_3cm$) structure below 1300 K, much higher than film deposition temperature.[9] Such a unique high-temperature low-symmetry structure allows charged point defects to migrate according to the site preference of point defects,[10-13] and thus $D_{defect}$ can be formed during film growth process. The internal bias developed by $D_{defect}$ provides a reasonable explanation for the double *P-E* loops.

To investigate the origin of the AFE-like double *P-E* loops, possible external parameters that can affect the physical properties of HMO thin films should be checked. First, we investigated the strain effect, which is the most common effect in epitaxial thin films. We fabricated two kinds of HMO thin films of different structures using a pulsed laser deposition (PLD) method: an epitaxial thin film experiencing a tensile in-plane strain[8,14] on the Pt(111)/Al$_2$O$_3$(0006) substrate and a polycrystalline thin film on the Pt(111)/SiO$_2$/Si(100) substrate. A sintered ceramic pellet of stoichiometric HMO was used as a target for the PLD. During growth, the oxygen partial pressure ($P_{O2}$) and the substrate temperature were maintained at 10 mTorr and 850 °C, respectively. Figures 1(b) and 1(c) show the X-ray pole figures for HMO ($11\bar{2}2$) diffraction peaks measured using a synchrotron radiation X-ray source. The randomly oriented HMO ($11\bar{2}2$) diffraction peaks in Fig. 1(c) show the polycrystalline growth on the Pt/SiO$_2$/Si substrate, while the epitaxial growth on the Pt/Al$_2$O$_3$ substrate is confirmed by the sixfold oriented HMO ($11\bar{2}2$) diffraction peaks in Fig. 1(b). Figures 1(a) and 1(d) show *P–E* loops for epitaxial and polycrystalline thin films, respectively, measured with a TF analyzer (aixACCT) at 2 kHz and 70 K. The C–V data, displayed in an inset in Fig. 1(d), also show a double butterfly feature, a hallmark of AFE-like behaviors. Therefore, the AFE-like double *P–E* loops could occur at a low temperature, irrespectively of the epitaxial strain. This indicates that the strain in the film cannot play an important role in the emergence of the AFE-like behavior.

Next, we considered defects in HMO thin films, since the experimental environment is always exposed to various sources of metal impurities, which have induced several abnormal experimental observations.[12,15-18] Actually, the proton-induced X-ray emission (PIXE) measurements using the reference samples of PACS-1 and $HoF_3$ showed traces of a small amount of metal impurities in HMO thin films, such as K, Ca, Ti, Cr, and Fe, whose concentration was less than 0.1 at. %. However, it was difficult to precisely determine the amount of each metal impurity in HMO thin films, due to the small thickness (~100 nm) of films and the strong signal from the substrate.

Electron paramagnetic resonance (EPR) is a powerful local probe technique used to study the defects (especially, substitutional ones).[19] We performed EPR measurements at 9.8 GHz using a Bruker EMX EPR spectrometer. Figure 2(a) shows the EPR spectra at $\theta = 0°$, 30°, 60°, and 90° for HMO thin films, where $\theta$ is the angle between the applied magnetic field ($H$-field) direction and the surface-normal direction of the film. The EPR peaks from HMO thin films are indicated by arrows, and extra peaks are from the $Al_2O_3$ substrate.[20] The strong resonance peak without a complex hyperfine structure indicates that a defect of nuclear spin $I = 0$ should exist. The hyperfine feature in EPR spectra was not observed even in the refined measurement of EPR (the inset of Fig. 2(a)).

Figure 2(b) shows the angular variation of the resonance field $H_{res}$ and the corresponding effective g-factor $g^{eff}$,[19] according to which $g_{\parallel}^{eff} = 1.986$ and $g_{\perp}^{eff} = 3.940$. Assuming that the defect with $I = 0$ is located at the Ho1 site having the $D_{3d}$ local symmetry[21] [Fig. 3(a) and 3(b)], the spin Hamiltonian can be written as

$$H = D_1(S_z^2 - 1/4) + D_2(S_z^2 - 1/4)(S_z^2 - 9/4) + g_\parallel \mu_B H \cos\theta + g_\perp \mu_B H \sin\theta, \quad (1)$$

where $D_1$ and $D_2$ are constants describing the zero-field crystal split in the D$_{3d}$ local environment. The assumption that the defect is located at the Ho1 site is justified later by theoretical calculations. Considering only the transition between $S_z = -1/2$ and $S_z = +1/2$,[19] this spin Hamiltonian gives the angular dependence[22] of

$$g^{eff} \cong [g_\parallel^2 + \{(S(S+1) + \frac{1}{4})g_\perp^2 - g_\parallel^2\}\sin^2\theta]^{1/2} \cdot [1 - \frac{1}{4}(S(S+1) - \frac{3}{4}) \cdot (g\beta H / 2D_1)^2]. \quad (2)$$

As shown in Fig. 2(b), the experimental data are well fitted by Eq. (2) with $S = 3/2$. Possible defects having $S = 3/2$ are $Cr^{3+}$, $Mn^{2+}$, and $Fe^{1+}$. [Other metal defects, such as K, Ca, and Ti detected by PIXE measurements, cannot have $S = 3/2$.] It should be noted that the nuclear spin of most isotopes of Fe (except the $^{57}$Fe of 2.1 % natural abundance) is $I = 0$. On the other hand, among the isotopes of Cr and Mn, the $^{53}$Cr of 9.5 % natural abundance and the $^{55}$Mn of 100 % natural abundance have the non-zero nuclear spin (i.e., $I \neq 0$), which would induce hyperfine features in their EPR spectra.[23] Thus, the result ($S = 3/2$, $I = 0$) of EPR measurements imply that the $Fe^{1+}$ should be the dominant substitutional defects in our HMO thin films.

Even if the $Fe^{1+}$ is uncommon compared to $Fe^{2+}$ and $Fe^{3+}$, the $Fe^{1+}$ state is probable[24] because it satisfies the local charge compensation together with one oxygen vacancy ($V_O$), which is also one of the most common defects in oxides. In other words, our EPR data implies that the acceptor-type $Fe^{1+}$ can form a $D_{defect}$ together with the donor-type $V_O$.

However, our EPR data doesn't provide direct information on the possible $D_{defect}$ ($Fe^{1+}$-$V_O$), but only on the predominance of the $Fe^{1+}$ defects in our films. Thus, to obtain further insight on the possible formation and configuration of $D_{defect}$, we performed first principles calculations based on the density functional theory for the HoMnO$_3$ ($\sqrt{3}\times\sqrt{3}\times2$) cell of the ferroelectric phase (*P*6$_3$*cm*). [In our calculations, it was quite difficult to treat the f-electrons of Ho properly. For the calculations, we replaced Ho by Y atoms, since YMnO$_3$ and HoMnO$_3$ have structural and ferroelectric properties that are quite similar to each other. Using the given experimental HoMnO$_3$ coordinates, we fully relaxed the YMnO$_3$ lattice structure as a reference. The details on the calculations are described elsewhere.[25]] The site of the Fe defect was determined to be the Ho1 site in Fig. 3(a) based on total energy calculations. The relative total energies for each site of Fe were 0, 0.02, and >10 eV at the Ho1, Ho2, and Mn sites, respectively. Our density of states (DOS) calculation also showed that a charge transfer from neighboring $V_O$ to Fe occurred. As a result of the charge transfer, the oxidation number of Fe became smaller than +3 and the predominance of the Fe$^{1+}$ defects observed by EPR measurements could be explained.

The pinning effect by $D_{defect}$ was also confirmed by the theoretical calculations. Figure 3(b) shows a local environment of Fe in which four sites are available for $V_O$. To determine the most probable site for $V_O$, the total energy was calculated for each site of $V_O$ in Fig. 3(b). The black solid circles in Fig. 3(c) clearly indicate that $V_O$ prefers site *B*. According to the Boltzmann factors (blue solid squares), the probability of $V_O$ occupancy at site *B* is larger by at least three orders of magnitude than at other sites. This result is also consistent with the recent reports that the distribution of point defects might follow the crystalline symmetry.[12,13] From this high preference, the Fe ion and $V_O$ complex should form an irreversible $D_{defect}$, which should give a strong internal bias to the ferroelectric polarization ($P_{FE}$) of HMO: the $P_{FE}$ direction should be pinned to be parallel to the $D_{defect}$ direction, as shown in Fig. 3(a, b).

The pinning of ferroelectric domains by irreversible $D_{defect}$ can induce the double *P*–*E*

loops. Usually, ferroelectric domains are formed by competition between the depolarization energy and the domain wall energy. Thus, on cooling after deposition, the HMO film enters the ferroelectric phase below $T_C$, with the formation of 180° domains. In the 180° domains, $D_{defect}$ might be aligned following the $P_{FE}$ direction through the $V_O$ migration. Considering the random distribution of $D_{defect}$, the resultant overall alignment of irreversible $D_{defect}$ in the film would be antiparallel-like, as shown in Fig. 4(a). If we apply an electric field to HMO thin films, the domains will be switched to the field direction, with the antiparallel-like alignment of irreversible $D_{defect}$ unchanged [Fig. 4(b)]. However, if the electric field is removed, the domain pattern will recover immediately into the original unpoled state [Fig. 4(c)] because the irreversible $D_{defect}$ can provide a strong local internal bias to recover the 180° domains.[12] The same result can be obtained for the reversed electric field [Fig. 4(d)]. Thus, the strong pinning effects by irreversible $D_{defect}$ can explain the double P-E loops of HMO thin films.[26]

For the HMO thin film which has a high Curie temperature, its high deposition temperature (850 °C) might allow the formation of irreversible $D_{defect}$ through the migration of charged point defects during film growth process. Also, compared to other ferroelectrics, the HMO has a highly distorted crystalline structure. The characteristic structure of HMO could induce a large energy difference among defect configurations (Fig. 3), much larger than that in other ferroelectrics.[27] The large energy difference would induce a significant internal bias by defect dipoles,[28] which can result in the double hysteresis, even for the small concentration (≤0.1 at. %) of Fe impurities. For $BaTiO_3$ ceramics, it has been reported that unintentionally-introduced Fe impurities of 0.02 at. % could induce a clear double hysteresis loop after aging process.[12]

The AFE-like behavior can be used to obtain a recoverable piezoelectric response to an external field[12,13] and a high tunability of the dielectric constant.[29] For these purposes, AFE-like behavior has previously been induced artificially, either by interfacial coupling using a layered thin-film structure[29] or by the redistribution of point defects using an additional aging process.[12,13] Our work demonstrates that the AFE-like double $P$-$E$ loops could be observed without artificial structure fabrication or any post-aging process.

In summary, we investigated the double $P$-$E$ loops in as-grown HMO thin films. Based on EPR measurements and first principles calculations, we found that the irreversible $D_{defect}$ in the film could be formed during film growth process by migration of point defects and could provide the domain pinning effect to cause the double $P$-$E$ loops in the as-grown thin film. We suggest the prospect of using the domain pinning effect of point defects to analyze and tune the physical properties of materials.

This work was supported by the Korea Science and Engineering Foundation (KOSEF) grant funded by the Korea government (MEST) (No. 2009-0080567). C.E.L. acknowledges support by National Research Laboratory R0A-2008-000-20066-0. The X-ray measurement was performed at the 10C1 beamline of Pohang Light Source. D.L. acknowledges support from a Seoul Science Scholarship.

<Figure captions>

FIG. 1. (Color online) $P$-$E$ loops measured at 70 K for HoMnO$_3$ thin films deposited on (a) Pt/Al$_2$O$_3$ and (d) Pt/TiO$_2$/SiO$_2$/Si substrates. The inset in (d) shows the capacitance vs electric field hysteresis loop of the HoMnO$_3$/Pt/TiO$_2$/SiO$_2$/Si thin film. Figures (b) and (c) are X-ray pole figures for (11$\bar{2}$2) diffraction peaks of HoMnO$_3$/Pt/Al$_2$O$_3$ and HoMnO$_3$/Pt/TiO$_2$/SiO$_2$/Si thin films, respectively.

FIG. 2. (Color online) (a) EPR spectra measured at $\theta$ = 0°, 30°, 60°, and 90°. The resonance peaks from the HoMnO$_3$ thin film are indicated by arrows. Extra peaks are from the Al$_2$O$_3$ substrate. The inset shows the EPR spectrum at $\theta$ = 0° by the refined measurement, in which the signal from the substrate is subtracted. (b) Angular variation of $H_{res}$ (black solid circles) in EPR spectra and $g^{eff}$ (red open circles) of the HoMnO$_3$ thin film. The red line is from Eq. (2) with S = 3/2.

FIG. 3. (Color online) (a) Crystalline structure of HoMnO$_3$ with a Fe defect. (b) Local environment of a Fe defect with four possible sites for an $V_O$, as indicated by the gray region in (a). (c) Relative values (black solid circles) of total energy for each site of the $V_O$ and corresponding Boltzmann factors (blue solid squares) calculated at $T_C$ = 875 K.

FIG. 4. (Color online) Schematic explanation of why the HoMnO$_3$ thin film with $D_{defect}$ can have double hysteresis responses. (a) As-grown state of the HoMnO$_3$ thin film. Upward (downward) triangles indicate the domains with +$c$ (−$c$)-oriented polarization. Red (blue) arrows indicate irreversible $D_{defect}$ along +$c$ (−$c$) axis. (b), (c), and (d) indicate the ferroelectric domain patterns when negative bias, zero bias, and positive bias, respectively, are applied sequentially.

Figure 1, D. Lee et al.

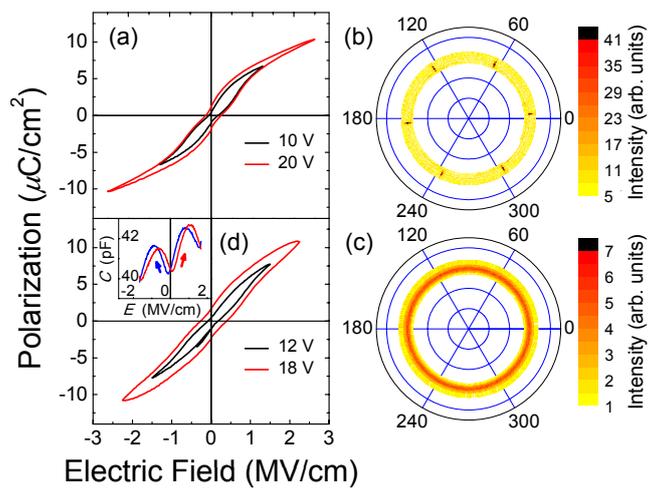

Figure 2, D. Lee et al.

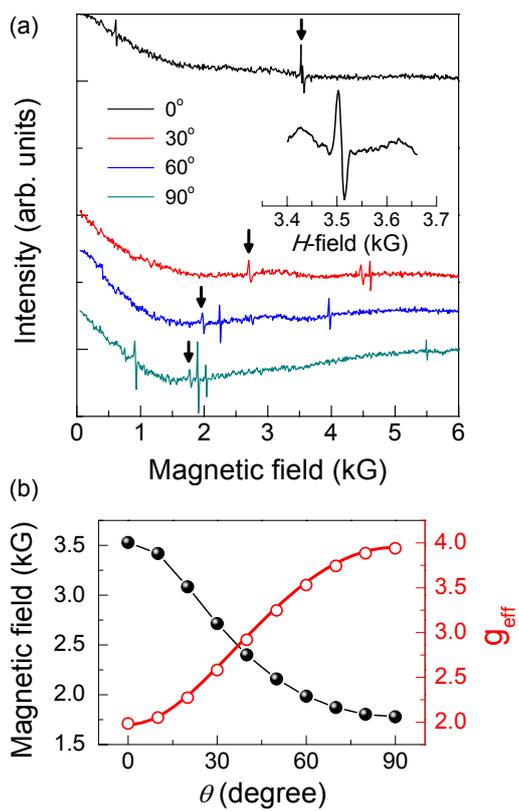

Figure 3, D. Lee et al.

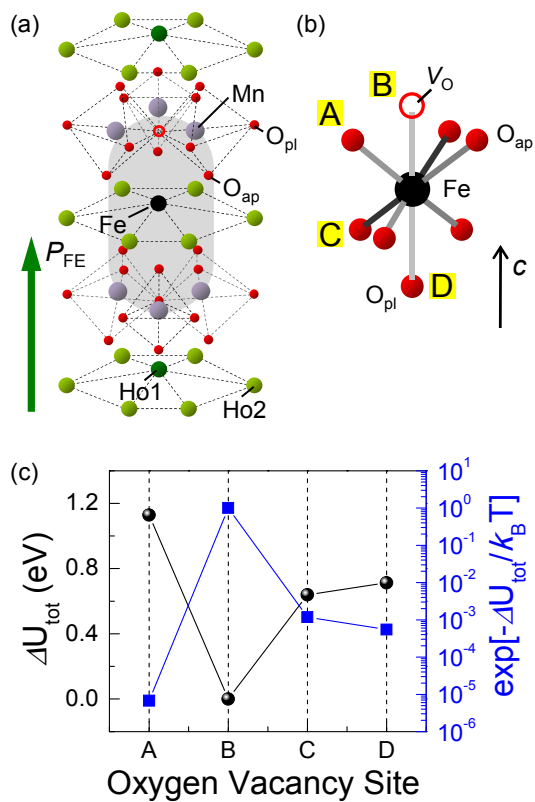

Figure 4, D. Lee et al.

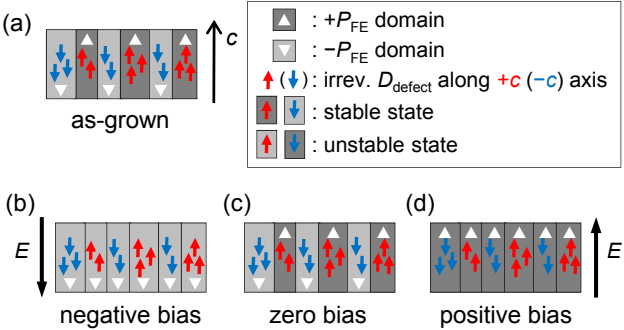